\documentclass[conference]{IEEEtran}

\usepackage{cite}
\usepackage{amsmath,amssymb,amsfonts}
\usepackage{algorithmic}
\usepackage{graphicx}
\usepackage{textcomp}
\usepackage{xcolor}
\usepackage{booktabs}
\usepackage{multirow}
\usepackage{float}
\usepackage{pifont}
\usepackage[table,xcdraw]{xcolor}
\usepackage{placeins}

\begin{document}

\title{StyGazeTalk: Learning Stylized Generation of Gaze and Head Dynamics}

\author{
Chengwei Shi, Chong Cao \\
Beihang University \\
{\tt\small shichengwei@buaa.edu.cn, chong@buaa.edu.cn}
}

\maketitle

\begin{abstract}
Gaze and head movements play a central role in expressive 3D media, human–agent interaction, and immersive communication. Existing works often model facial components in isolation and lack mechanisms for generating personalized, style-aware gaze behaviors. We propose StyGazeTalk, a multimodal framework that synthesizes synchronized gaze–head dynamics with controllable styles. To support high-fidelity training, we construct HAGE, a high-precision multimodal dataset containing eye-tracking data, audio, head pose, and 3D facial parameters. Experiments show that our method produces temporally coherent, style-consistent gaze–head motions, enhancing realism in 3D face generation. Code will be released upon acceptance.
\end{abstract}


\section{Introduction}
\label{sec:intro}

Realistic 3D facial animation increasingly relies on expressive gaze--head behavior to convey attention, intention, and social presence \cite{Yu2012}. However, most existing systems prioritize lip synchronization or lower-face motion, without explicitly modeling gaze dynamics. Rule-based gaze systems \cite{Pelachaud2003, Koda2017, Ruhland2014, Lee2002} are interpretable but fail to capture the temporal richness and person-specific variability of natural gaze, while data-driven approaches often model facial attributes independently \cite{xing2023codetalker, VOCA2019, faceformer2022, wang2021audio2head, makeittalk}, rely on monocular estimation \cite{sun2024diffposetalk, DECA:Siggraph2021}, or encode style using discrete speaker identities \cite{xing2023codetalker, faceformer2022}, resulting in inconsistent gaze--head coordination and limited stylistic expressiveness.We address these gaps by explicitly treating gaze as a first-class dynamic modality: we construct a high-quality multimodal dataset with synchronized audio--video and calibrated gaze--head data captured by a professional eye tracker, explicitly model the temporal and physiological regularities of human gaze to generate coordinated eye--head dynamics, and introduce a continuous style encoder with pattern-level metrics for realism evaluation.

\begin{figure}[t]
    \centering
    \includegraphics[width=\linewidth]{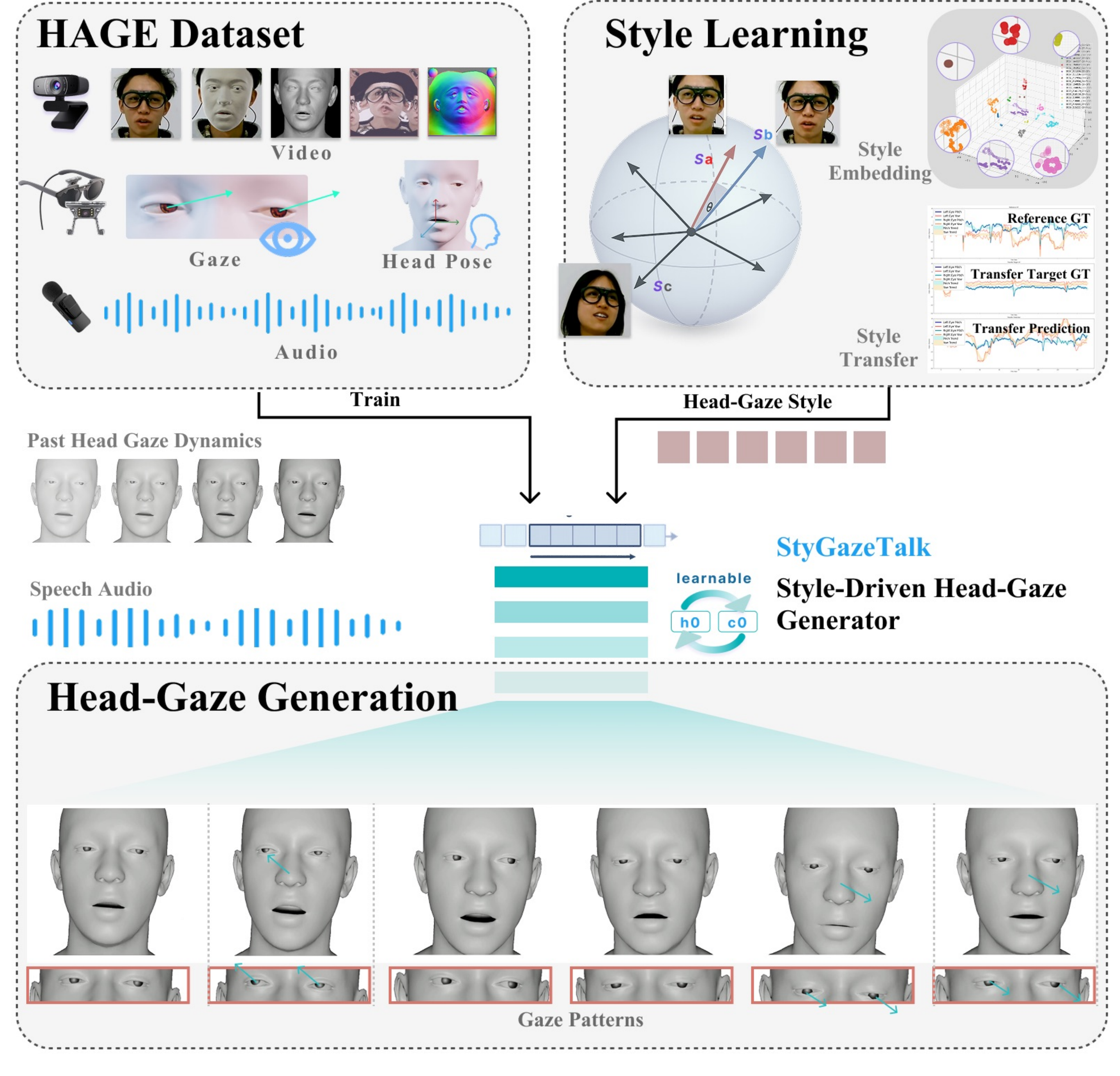}
    \caption{Overview of the proposed StyGazeTalk framework.}
    \label{fig:teaser}
\end{figure}

As shown in Fig.~\ref{fig:teaser}, our main contributions are:

\begin{itemize}
    \item \textbf{A physiologically grounded approach for generating coordinated gaze--head dynamics} (StyGazeTalk), capturing cross-modal temporal dependencies and characteristic gaze--head patterns.

  \item \textbf{A high-precision multimodal dataset} (HAGE) with synchronized audio, expression, gaze, and head pose collected via professional eye tracking in unscripted conversations, complemented by pattern-level evaluation metrics grounded in eye-movement research.

  \item \textbf{A learned style encoder} that learns a continuous space of gaze–head style and enables controllable style generation.
\end{itemize}

\begin{figure*}[t] 
    \vspace{-.25em}
    \centering
    \includegraphics[width=\textwidth]{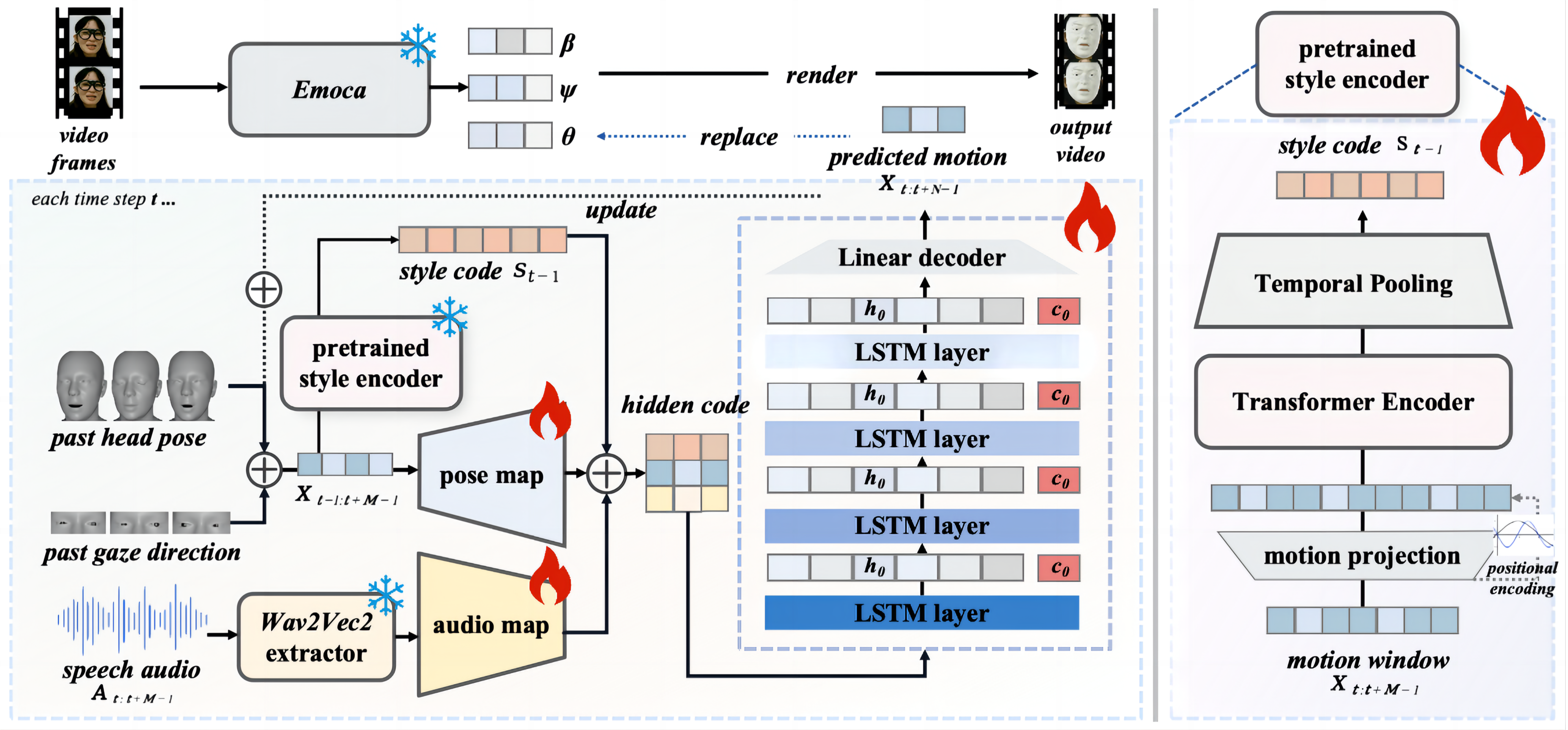}
    \vspace{-.25em}
    \caption{ Overview of our method. (Left) At each time step t, the model takes audio $A_{t:t+M}$, past motion $X_{t-1:t+M}$, and a style code $s_{t-1}$, and predicts future motion $X_{t:t+N-1}$ using a conditional sequence generation model. (Right) The style code is extracted from past motion via a pretrained Transformer encoder, enabling style-aware, temporally coherent motion generation.}
    \label{fig:overview}
\end{figure*}

\section{Related Work}
\label{sec:Related Work}

\subsection{Speech-Driven 3D Facial Animation}
Parametric 3DMMs (e.g., FLAME~\cite{FLAME:SiggraphAsia2017}) enable controllable facial animation and support systems such as CodeTalker~\cite{xing2023codetalker}, VOCA~\cite{VOCA2019}, and MeshTalk~\cite{richard2021meshtalk}. With pretrained audio encoders like Wav2Vec2~\cite{baevski2020wav2vec20frameworkselfsupervised}, these methods achieve strong lip sync but primarily model lower-face motion and treat modalities separately, leaving gaze and head dynamics under-modeled.

\subsection{Head Pose Generation}
Head motion generation has been explored via 2D keypoint/flow manipulation~\cite{makeittalk, wang2021audio2head, zhang2022sadtalker} and 3D regression~\cite{sun2024diffposetalk, faceformer2022}. Although 3D approaches offer better pose consistency, most depend on monocular estimators~\cite{openface, lugaresi2019mediapipeframeworkbuildingperception, spectre, DECA:Siggraph2021} derived from VoxCeleb~\cite{VoxCeleb}, HDTF~\cite{zhang2022sadtalker}, etc., which introduce noise, drift, and expression entanglement—limiting realism.

\subsection{Style Modeling}
Style is commonly encoded using one-hot speaker IDs~\cite{chung2019voice2face, li2019fsgan, sahipjohn2024dubwisevideoguidedspeechduration, gururani2022space, Imitator, faceformer2022, wav2lip, zhang2022sadtalker, VOCA2019}, which ignores inter-/intra-speaker variability. Recent contrastive approaches~\cite{contrastivelearning1, contrastivelearning2, contrastivelearning3} instead learn continuous embeddings; following this direction, we adopt a contrastive style encoder tailored to gaze–head dynamics.

\section{Method}
\label{sec:method}

Figure~\ref{fig:overview} summarizes our framework for generating temporally coherent and style-consistent 3D gaze–head motion from speech, explicitly modeling the structured regularities of gaze dynamics.

\subsection{Pattern-Informed Generation}
\begin{figure}[t]

    \centering
    \includegraphics[width=.5\textwidth]{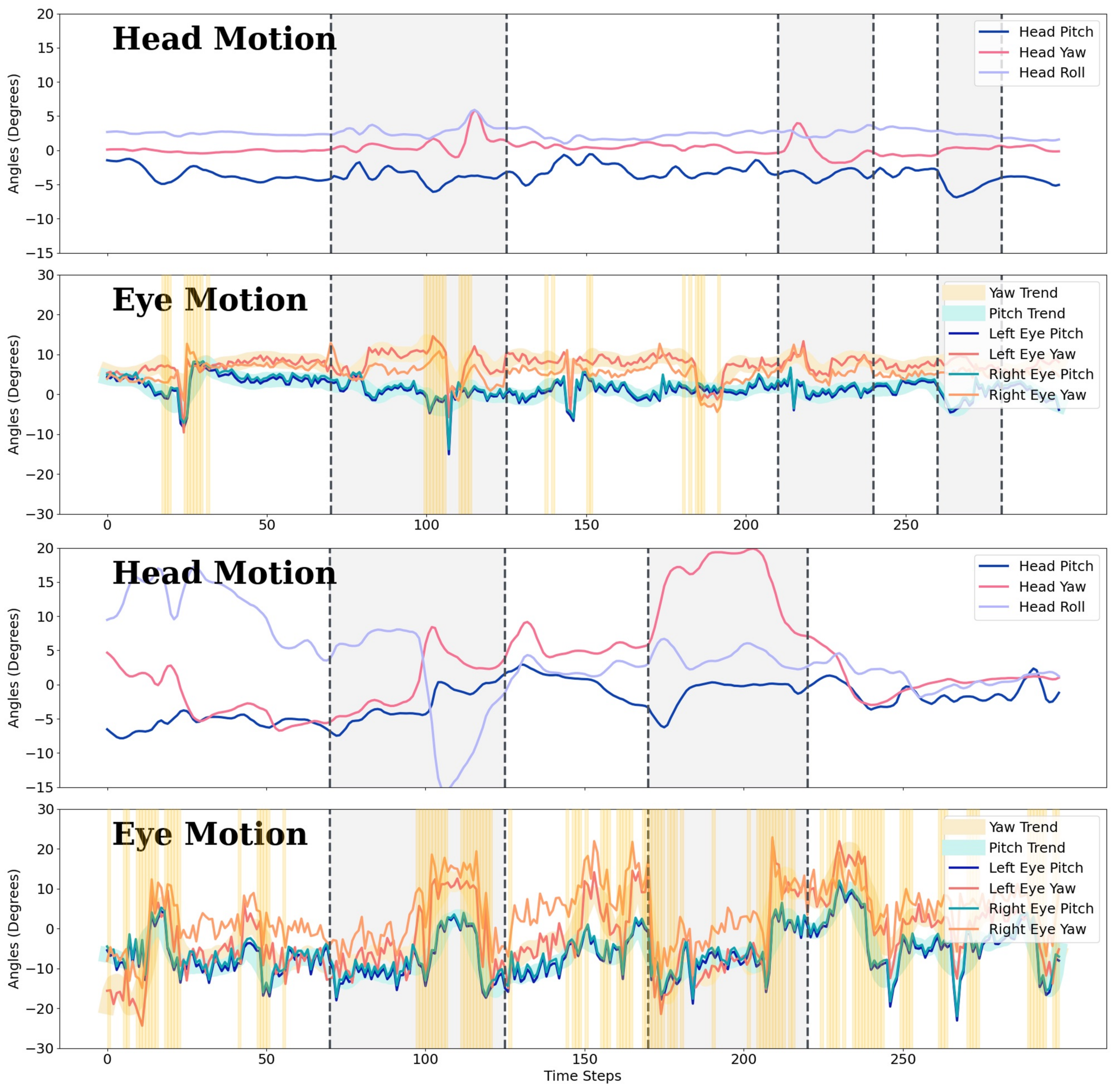} 

    \caption{Structured gaze–head dynamics: fixations/saccades (yellow), eye–head coordination (gray), and speaker-specific style patterns.}
    \label{fig:pattern}
\end{figure}

Unlike arbitrary signals, gaze behavior follows structured physiological dynamics. Building on findings in eye-movement research~\cite{pattern-idt, pattern1, pattern2, head-gaze-pattern, head-gaze-pattern2, threshold1, threshold2}, our data (Fig.~\ref{fig:pattern}) reveals two characteristic gaze–head patterns, which motivate our subsequent modeling design.

(1) \textbf{Shared patterns} capture universal gaze–head dynamics, including long-term gaze–head compensation and short-term fixation/saccade cycles~\cite{threshold1, threshold2}.

(2) \textbf{Sample-specific patterns} capture speaker-consistent motion tendencies within local windows, such as characteristic velocity and amplitude profiles.

To model these pattern types, we use a multi-layer LSTM that captures both short- and long-term gaze–head dynamics. Its learnable initial states encode global temporal priors, while the sliding-window formulation captures segment-level variations and naturally supports streamable generation when future context is unavailable. A style encoder extracts locally consistent cues from preceding motion. We adopt this architecture over Transformer-based alternatives, as it provides more stable optimization under our data scale and supports causal windowed generation ~\cite{LSTMVSTRANS}; ablations (Sec.~5 and supplementary material) validate this choice.

\subsection{Problem Formulation}
Let $\mathbf{A}_{1:T} = (\mathbf{a}_t)$ denote the audio feature sequence and $\mathbf{X}_{1:T} = (\mathbf{x}_t)$ the corresponding 7-DoF motion (head pitch/yaw/roll and binocular gaze angles). Our objective is to learn a speech-conditioned function that predicts future motion with style control. At time $t$, the model receives an audio window $\mathbf{A}_{t:t+M-1}$, the previous motion window $\mathbf{X}_{t-1:t+M-1}$, and a style embedding $\mathbf{s}_{t-1}$ extracted from the preceding motion. It outputs the next motion segment $\mathbf{X}_{t:t+N-1}$.

\subsection{Synthesis of Gaze-Pose Dynamics}

We extract contextualized audio features using Wav2Vec~2.0~\cite{baevski2020wav2vec20frameworkselfsupervised}.   
For each step $t$, an audio window $\mathbf{A}_{t:t+M-1}$ from the raw waveform is encoded and projected into a $d$-dimensional space $\tilde{\mathbf{A}}_{t:t+M-1}$, while the previous motion window $\mathbf{X}_{t-1:t+M-1}$ is mapped to $d$-dimension as $\tilde{\mathbf{X}}_{t-1:t+M-1}$.  
We adopt a two-stage scheme: a contrastive style encoder (SE) is first pretrained following~\cite{contrastivelearning1, contrastivelearning2, contrastivelearning3} to learn a continuous style embedding, producing
\[
\mathbf{s}_{t-1} = \mathrm{SE}(\mathbf{X}_{t-1:t+M-1}),
\]
which captures stable gaze–head dynamic style within the local window.  
We concatenate audio features, motion features, and the style vector to form the generator input  
$\mathbf{Z}_t \in \mathbb{R}^{M \times (2d + d_s)}$. A multi-layer LSTM generator (MG) then processes $\mathbf{Z}_t$, with learnable initial states $(\mathbf{h}_0,\mathbf{c}_0)$ encoding shared temporal priors and ensuring coherence across windows, to predict the next motion segment:
\[
\hat{\mathbf{X}}_{t:t+N-1} = \mathrm{MG}(\mathbf{Z}_t; \mathbf{h}_0,\mathbf{c}_0).
\]

While our model predicts only dynamic gaze and head motion, we optionally fuse these outputs with shape, jaw, and expression parameters extracted by EMOCA~\cite{EMOCA} at inference for visualization. This fusion is used solely to render complete facial animations and does not affect training or prediction.

We train the motion generator with position and velocity supervision.  
The frame-wise term is an MSE loss:
$\mathcal{L}_{\text{mse}}=\tfrac{1}{T}\sum_{t=1}^{T}\|\hat{\mathbf{x}}_t-\mathbf{x}_t\|_2^2$
and motion smoothness is enforced by a velocity loss:
$\mathcal{L}_{\text{vel}}=\tfrac{1}{T-1}\sum_{t=2}^{T}\|(\hat{\mathbf{x}}_t-\hat{\mathbf{x}}_{t-1})-(\mathbf{x}_t-\mathbf{x}_{t-1})\|_2^2.$
The generator is trained with $\mathcal{L}_{\text{gen}}=\lambda\,\mathcal{L}_{\text{mse}}+(1-\lambda)\,\mathcal{L}_{\text{vel}}$.

For the style encoder, we adopt an NT-Xent contrastive objective~\cite{contrastivelearning1}.  
Given style embeddings  
$\mathcal{S}=\{\mathbf{s}_1^a,\ldots,\mathbf{s}_N^a,\mathbf{s}_1^b,\ldots,\mathbf{s}_N^b\}\in\mathbb{R}^{2N\times d_s}$  
and cosine similarity  
$\mathrm{sim}(\mathbf{s}_i,\mathbf{s}_j)=\frac{\mathbf{s}_i^\top\mathbf{s}_j}{\|\mathbf{s}_i\|\|\mathbf{s}_j\|}$,  
the loss is
\[
\mathcal{L}_{\text{con}}
=-\sum_{i=1}^{2N}\log
\frac{\exp(\mathrm{sim}(\mathbf{s}_i,\mathbf{s}_{p(i)})/\tau)}
{\sum_{k\neq i}\exp(\mathrm{sim}(\mathbf{s}_i,\mathbf{s}_k)/\tau)} ,
\]
where $p(i)$ is the positive pair and $\tau$ is a temperature term.  
This encourages style consistency within a temporal motion segment, yielding stable style embeddings that can be exploited for generation.

\section{HAGE Dataset}

The Head-Pose–Audio–Gaze–Expression (HAGE) dataset is a high-fidelity multimodal corpus designed specifically for learning physiologically realistic gaze–head coordination. It contains densely annotated, unscripted conversational recordings collected with a professional binocular eye tracker, providing synchronized 16 kHz audio, 1080×1080 video at 25 FPS, and sensor-accurate binocular gaze and 3D head rotations.
Unlike video-derived datasets that rely on noisy monocular estimation, HAGE offers drift-free, high-precision gaze supervision, preserving the fine-grained temporal structure required to model fixations, saccades, and gaze–head compensation.

\begin{table}[htbp]
\centering
\setlength{\tabcolsep}{3.5pt}  
\renewcommand{\arraystretch}{1.1}

\caption{Multimodal Dataset Comparison.}
\label{tab:datasets}

\begin{tabular}{l|c|c|c|c|c|c|c}
\hline
\textbf{Dataset} &
\textbf{Audio} &
\textbf{Video} &
\textbf{HP} &
\textbf{Gaze} &
\textbf{Exp} &
\textbf{Acquisition} &
\textbf{Content} \\
\hline

VOCASET    & \ding{51} & \ding{51} & \ding{51} & \ding{55} & \ding{51} & 4D Scans    & Scripted \\
MEAD       & \ding{51} & \ding{51} & \ding{55} & \ding{55} & \ding{51} & RGB         & Scripted \\
BIWI       & \ding{51} & \ding{51} & \ding{51} & \ding{55} & \ding{55} & RGB-D       & None \\
TFHP       & \ding{51} & \ding{51} & \ding{51} & \ding{55} & \ding{51} & RGB         & Formal Video \\
HAGE & \ding{51} & \ding{51} & \ding{51} & \ding{51} & \ding{51} & Eye Tracker & Unscripted \\
\hline
\end{tabular}
\end{table}

\textbf{Comparison of Datasets.}
Table~\ref{tab:datasets} summarizes existing talking-face datasets (\textit{HP = Head Pose, Exp = Expression}). Most lack gaze data or rely on scripted speech or vision-based estimation. Our HAGE dataset uniquely offers sensor-accurate gaze and head motion from unscripted conversations, which naturally elicit spontaneous attention shifts and natural gaze–head coordination\cite{unscriped}. Despite its compact duration, HAGE provides sensor-level gaze supervision that is substantially more informative than video-estimated signals, and shows consistent performance across all our evaluations.

\subsection{Gaze Pattern Metrics}
We evaluate gaze–head realism using three pattern-level metrics—Fixation Ratio, Compensation Score, and Similarity with Ground Truth—following established findings in eye-movement research.
\subsubsection{Fixation Ratio}
Fixations are detected using the standard dispersion-based I-DT algorithm~\cite{pattern-idt}. We follow common practice in eye-movement research and apply the typical angular-dispersion and minimum-duration criteria used in prior work~\cite{fixation_threshold}. The fixation ratio is then computed as $\text{FixRatio}=N_{\text{fix}}/(N_{\text{fix}}+N_{\text{sac}})$.

\begin{table*}[t]
\centering
\caption{Quantitative evaluation of our full gaze–head models and ablations.}
\label{tab:full}
\resizebox{\textwidth}{!}{
\renewcommand{\arraystretch}{1.4}
\begin{tabular}{c|ccccc|cccc|cccc}
\hline
\textbf{Method} 
& \multicolumn{5}{c|}{\textbf{All}} 
& \multicolumn{4}{c|}{\textbf{Gaze}} 
& \multicolumn{4}{c}{\textbf{Head}} \\
\hline
\textbf{Metric} 
& MAE↓ & Vel↓ & MEE↓ & CE↓ & BAS↑
& MAE↓ & Vel↓ & MEE↓ & BAS↑
& MAE↓ & Vel↓ & MEE↓ & BAS↑ \\
\hline
Base         & 5.149 & 6.840 & 57.350 & 0.139 & 0.219 & 6.616 & 11.888 & 57.174 & 0.205 & 3.193 & 0.108 & 0.398 & 0.211 \\
SE-32        & 5.097 & 6.863 & 56.836 & 0.138 & 0.221 & \textbf{6.527} & 11.928 & 56.655 & 0.231 & 3.189 & 0.109 & 0.403 & 0.207 \\
SE-64        & \textbf{5.087} & 6.843 & \textbf{56.411} & \textbf{0.136} & 0.236 
             & 6.535 & 11.894 & \textbf{56.215} & 0.238 
             & \textbf{3.156} & \textbf{0.107} & \textbf{0.390} & 0.205 \\
SE-64-VEL    & 5.219 & \textbf{6.715} & 56.616 & 0.152 & \textbf{0.243}
             & 6.645 & \textbf{11.668} & 56.445 & \textbf{0.240}
             & 3.317 & 0.112 & 0.413 & \textbf{0.212} \\
TFM-SE-64    & 5.659 & 7.852 & 71.177 & 0.137 & 0.197 & 7.353 & 13.618 & 70.938 & 0.199 & 3.400 & 0.166 & 0.611 & 0.203 \\
Pseudo-SE-64         & 6.004 & 7.013 & 58.023 & 0.148 & 0.203 & 7.601 & 12.165 & 56.71 & 0.202 & 3.873 & 0.114 & 0.421 & 0.201 \\
\hline
\end{tabular}}
\end{table*}

\begin{table}[t]
\centering
\caption{Comparison with ablation variants and head-only baselines.}
\label{tab:headonly}
\resizebox{0.47\textwidth}{!}{
\renewcommand{\arraystretch}{1.25}
\begin{tabular}{c|l|cccc}
\hline
Section & Method & MAE↓ & Vel↓ & MEE↓ & BAS↑ \\
\hline
\multirow{2}{*}{Ours} 
& SE-64     & 3.156 & 0.107 & 0.390 & 0.205 \\
& SE-64-VEL & 3.317 & 0.112 & 0.413 & 0.212 \\
\hline
\multirow{5}{*}{Head-only}
& SadTalker      & 3.935 & 0.173 & 0.618 & 0.199 \\
& DiffPoseTalk   & 4.415 & 0.093 & 0.327 & 0.199 \\
& Audio2Head     & 5.810 & 0.175 & 0.648 & 0.193 \\
& PC-AVS         & 3.937 & 0.199 & 0.719 & 0.197 \\
& MakeItTalk     & 3.939 & 0.211 & 0.766 & 0.198 \\
\hline
\end{tabular}}
\end{table}

\subsubsection{Compensation Score}
Following~\cite{head-gaze-pattern}, eye–head coordination is quantified using head velocity $\bar{\mathbf{h}}_t$ and mean eye velocity $\bar{\mathbf{e}}_t$. Net gaze shift is $\mathbf{g}_t=\bar{\mathbf{e}}_t+\bar{\mathbf{h}}_t$, and the per-frame score is  
$\text{Score}_t=\begin{cases}
-\|\bar{\mathbf{h}}_t\|, & \|\mathbf{g}_t\|<20,\\
-\cos(\theta), & 25\le\|\mathbf{g}_t\|\le90,\\
0, & \text{otherwise},
\end{cases}$  
where $\theta$ is the angle between $\bar{\mathbf{h}}_t$ and $\bar{\mathbf{e}}_t$. The final score is $\text{CompScore}=\frac{1}{T}\sum_t \text{Score}_t$.

\subsubsection{Similarity with Ground Truth}
Overall realism is assessed by  
$\text{Sim}=1-\big(|\text{Fix}_{\text{GT}}-\text{Fix}_{\text{pred}}|+|\text{Comp}_{\text{GT}}-\text{Comp}_{\text{pred}}|\big)$,  
capturing alignment with human fixation patterns and gaze–head coordination.

\section{Experiments}

\begin{figure*}[t] 
    \vspace{-.25em}
    \centering
    \includegraphics[width=.9\textwidth]{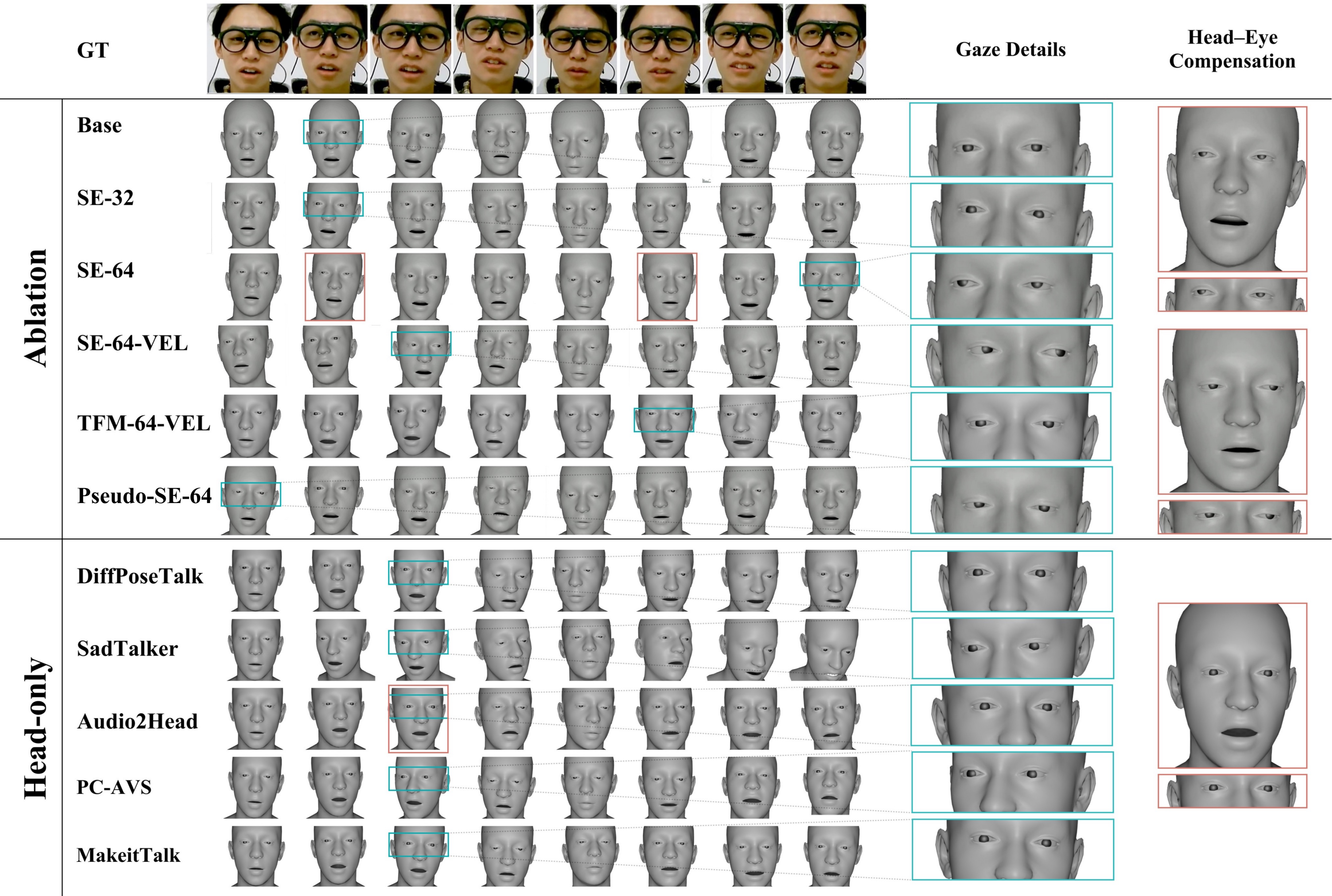} 
    \vspace{-.25em}
    \caption{ Visual results of ablation study of our methods and qualitative comparison of state-of-the-art methods.}
    \label{fig:compare}
\end{figure*}

\subsection{General Evaluation}
As shown in Table~\ref{tab:full} and Table~\ref{tab:headonly}, we evaluate our method through ablations and baseline comparisons on the HAGE dataset. Ablations include: the base LSTM model (\textit{Base}); style encoders of 32/64 dimensions (\textit{SE-32}/\textit{SE-64}); a version trained with velocity loss (\textit{SE-64-VEL}); a Transformer-based generator (\textit{TFM-SE-64}) with matched architecture and hyperparameters, together with standard Transformer-oriented tuning; and a variant trained with video-estimated gaze supervision instead of eye-tracking ground truth (\textit{Pseudo-SE-64}), where gaze is obtained using a state-of-the-art appearance-based estimator \cite{mpg}. For head-motion baselines—\textit{SadTalker} \cite{zhang2022sadtalker}, \textit{DiffPoseTalk} \cite{sun2024diffposetalk}, \textit{Audio2Head} \cite{wang2021audio2head}, \textit{PC-AVS} \cite{PC-AVS}, and \textit{MakeItTalk} \cite{makeittalk}—we report head-only metrics since they do not generate gaze. All methods are evaluated on HAGE to ensure comparable supervision conditions. We use five metrics: MAE, velocity error (Vel), motion energy error (MEE), and Beat Alignment Score (BAS) \cite{BA}, plus their gaze counterparts, covering frame accuracy (MAE), motion dynamics (Vel/MEE), and audio–motion synchrony (BAS). These metrics collectively reflect perceptual attributes: MAE aligns with directional accuracy, Vel and MEE capture smoothness and motion energy, and BAS correlates with perceived audio–motion timing.

\textit{SE-64} achieves the best overall performance, showing the benefit of style modeling, and adding velocity loss (\textit{SE-64-VEL}) further improves Vel and BAS, sharpening motion dynamics and audio alignment. \textit{TFM-SE-64} and \textit{Pseudo-SE-64} both degrade performance—especially for gaze—revealing the limits of Transformer backbones and pseudo-supervised gaze estimation, as well as the irreplaceable value of high-fidelity eye-tracking supervision. Compared with head-only baselines, our \textit{SE-64} variants still achieve lower head MAE and higher BAS while also generating gaze, whereas \textit{DiffPoseTalk} offers strong head motion but no coordinated eye–head behavior.

Qualitative results (Fig.~\ref{fig:compare}) show that our full model produces more coherent eye–head dynamics than both ablations and head-only baselines (set gaze to a zero direction). In contrast, our style-aware variants (\textit{SE-32}/\textit{SE-64}) enhance naturalness, the velocity term boosts motion variability, and Transformer/pseudo-supervised alternatives appear over-smoothed or unstable—supporting the effectiveness of our physiologically grounded design. More visual and temporal results, including tests on video-derived inputs, are provided in the supplementary video.

\subsection{Gaze Pattern Evaluation}

\begin{table}[t]
\centering
\caption{Results of gaze pattern evaluation.}
\label{tab:gaze_pattern_evaluation}

\resizebox{\linewidth}{!}{%
\begin{tabular}{lcccc}
\toprule
Section & saccades & fixation & compScore & simScore \\
\midrule
GT          & 38.87\% & 61.13\% & -0.3366 & 1.0000 \\
Base        & 25.83\% & 74.17\% & -0.2890 & 0.8220 \\
SE-32       & 29.02\% & 70.98\% & -0.3021 & 0.8670 \\
SE-64       & 30.78\% & 69.22\% & -0.3054 & 0.8879 \\
SE-64-vel   & 32.67\% & 67.33\% & -0.3100 & 0.9114 \\
TFM-SE-64   & 15.43\% & 84.57\% & -0.2006 & 0.6296 \\
Pseudo-SE-64        & 26.33\% & 73.67\% & -0.2873 & 0.7082 \\
\bottomrule
\end{tabular}}
\end{table}

Table~\ref{tab:gaze_pattern_evaluation} reports fixation ratio, compensation score, and the composite SimScore. 
Ground-truth gaze shows a 61.13\% fixation ratio and a compensation score of $-0.3366$, which is consistent with typical conversational gaze dynamics~\cite{fixation_threshold}. 
\textit{Base} over-predicts fixations (74.17\%) and weakens eye–head coordination. 
Style supervision improves both aspects: \textit{SE-32} reaches 0.8670 and \textit{SE-64} reaches 0.8879. \textit{SE-64-VEL} best matches real gaze statistics. 
\textit{TFM-SE-64} collapses into over-smoothed gaze, and \textit{Pseudo-SE-64} also underperforms with reduced saccades and weaker compensation. 
These trends show that high-frequency gaze dynamics cannot be recovered from video-only supervision or generic Transformer backbones, highlighting the necessity of our high-fidelity eye-tracking dataset.

\subsection{Style Encoder Evaluation}

We assess the style encoder through three tasks: style-metric scoring, clustering, and style transfer.

\subsubsection{Style Encoder Metrics}
Style consistency is evaluated by cosine distance between predicted and ground-truth style embeddings:
$\mathcal{E}_{\text{style}} = 1 - \cos(\mathbf{s}^{\text{gt}}, \hat{\mathbf{s}})$.
As shown in Table~\ref{tab:full}, \textit{Base}, \textit{SE-32}, and \textit{SE-64} obtain errors of 0.139, 0.138, and 0.136, respectively, while \textit{Pseudo-SE-64} yields the largest error (0.148), indicating the necessity of eye-tracking supervision.

\subsubsection{Clustering Visualization}
Figure~\ref{fig:gaze_motor_dynamics} shows t-SNE embeddings across 10 sessions. 
Ground-truth samples form clear session-wise clusters, and predictions align with them, indicating that the encoder captures stable session-specific dynamics.

\begin{figure}[t]
    \centering
    \includegraphics[width=0.45\textwidth]{fig/cluster.pdf}
    \caption{t-SNE of style embeddings. Colors indicate sessions; solid = ground truth, transparent = predictions.}
    \label{fig:gaze_motor_dynamics}
\end{figure}

\begin{figure}[t]
    \centering
    \includegraphics[width=0.45\textwidth]{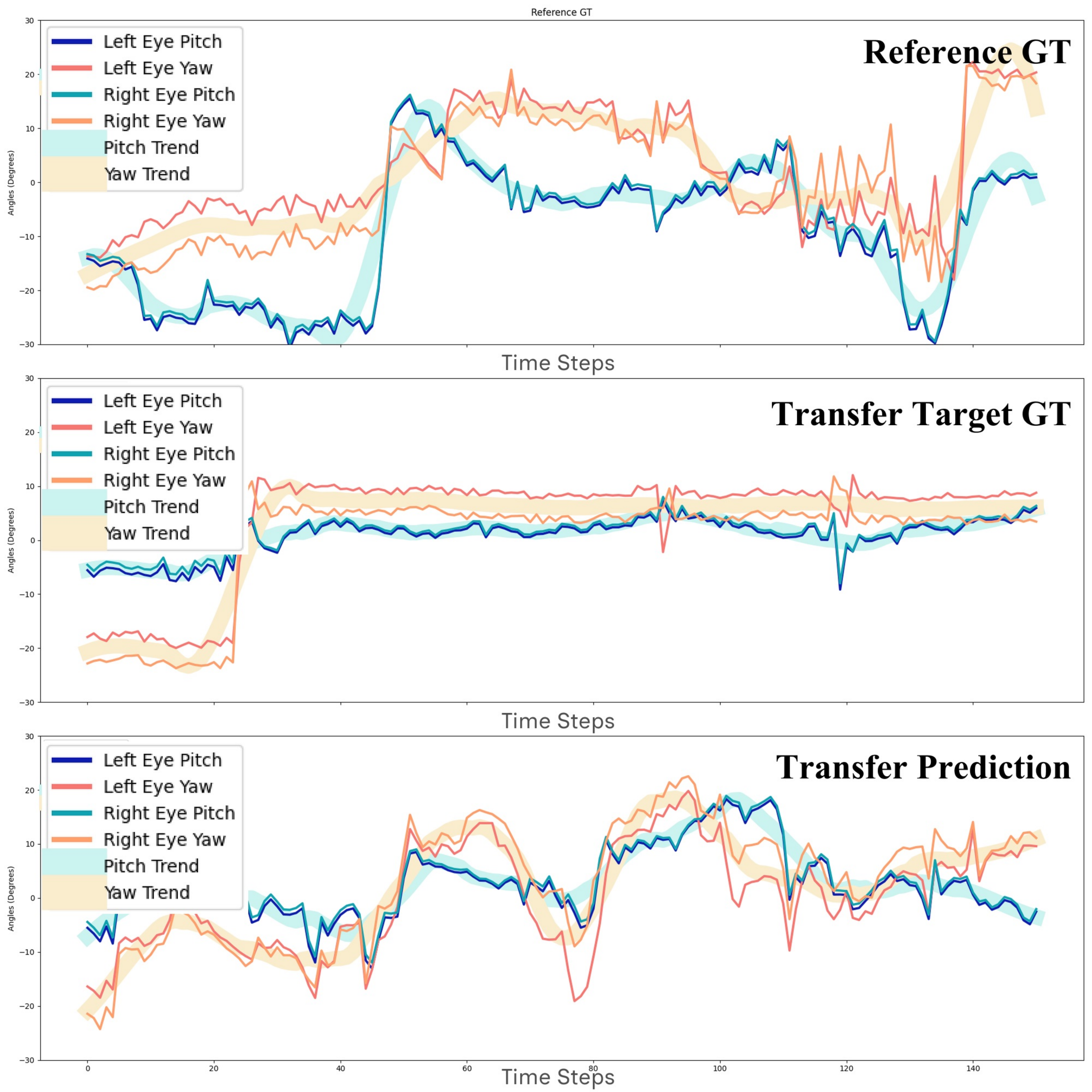}
    \caption{Style transfer results: reference style applied to target structure.}
    \label{fig:style_trans}
\end{figure}

\subsubsection{Style Transfer}
As shown in Fig.~\ref{fig:style_trans}, the model transfers temporal style while preserving target structure, indicating effective style–content disentanglement. Additional results are shown in the supplementary video.

\begin{table}[t]
\centering
\caption{User preference (\%) across dimensions.}
\label{tab:user_study_results}
\renewcommand{\arraystretch}{1.2}
\resizebox{\linewidth}{!}{%
\begin{tabular}{lccc}
\toprule
\textbf{Method} & \textbf{Naturalness} & \textbf{Realism} & \textbf{Style Similarity} \\
\midrule
DiffPoseTalk    & 24.5\% & 26.2\% & 24.4\% \\
TFM-SE-64       & 27.2\% & 25.3\% & 27.6\% \\
SE-64 (Ours)    & \textbf{48.3\%} & \textbf{48.5\%} & \textbf{47.9\%} \\
\bottomrule
\end{tabular}
}
\end{table}

\subsection{User Perception Study}

We conducted a forced-choice user study comparing \textit{SE-64}, \textit{TFM-SE-64}, and \textit{DiffPoseTalk}. 
Participants (78 valid) evaluated 12 randomly selected test videos on \textit{Naturalness}, \textit{Realism}, and \textit{Style Similarity} following prior protocols~\cite{sun2024diffposetalk}. 
As shown in Table~\ref{tab:user_study_results}, \textit{SE-64} is preferred in all dimensions. 
Differences are statistically significant (paired t-test, $p<0.05$); see supplementary for details.

\section{Conclusion}

We presented StyGazeTalk, a speech-driven gaze–head generation framework enabled by high-fidelity eye-tracking supervision. Experiments across objective, perceptual, and pattern-level evaluations show that accurate gaze sensing is critical for producing realistic and coordinated eye–head motion, suggesting its potential as a reliable supervisory signal for cross-domain or distillation-based learning.

\bibliographystyle{IEEEbib}
\bibliography{main}

\vspace{12pt}

\end{document}